\documentstyle[11pt]{article}
\def\Journal#1#2#3#4{{\em #1} {\bf #2}, #3 (#4)}
\def\Book#1#2#3{{\em #1} ({#2}, {#3})}
\def\Preprint#1#2#3#4{{\em #1}, preprint~{#2}, {#3} (#4)}
\def\JPA{J. Phys. A: Math. Gen.}
\begin{document}
\newcommand{\be}{\begin{equation}}
\newcommand{\ee}{\end{equation}}
\newcommand{\bea}{\begin{eqnarray}}
\newcommand{\eea}{\end{eqnarray}}
\newcommand{\beaa}{\begin{eqnarray*}}
\newcommand{\eeaa}{\end{eqnarray*}}
\newcommand{\qd}{\quad}
\newcommand{\qqd}{\qquad}
\newcommand{\npb}{\nopagebreak[1]}
\newcommand{\nn}{\nonumber}
\newcommand{\dbar}{\bar{\partial}}
%\begin{titlepage}
\title{\bf Analytic-bilinear approach to integrable hierarchies.\\
I.Generalized KP hierarchy.}
\author{L.V. Bogdanov\thanks{Permanent address: IINS,
Landau Institute for Theoretical Physics, Kosygin str. 2,
Moscow 117940, GSP1, Russia; e-mail Leonid@landau.ac.ru}
\hspace{0.1em} and B.G. Konopelchenko\thanks
{Also: Budker Institute of Nuclear Physics, Novosibirsk 90,
Russia}\\
Consortium EINSTEIN\thanks{European Institute for
Nonlinear Studies via Transnationally Extended Interchanges},
\\ Dipartimento di Fisica dell'Universit\`a
and Sezione INFN, \\73100 Lecce, Italy
%\\ \em International Institute for Nonlinear Sciences,\\ \em
%GSP-1 117940, 2 Kosygina, Moscow v-334, Russia
}
\date{}
\maketitle
%\end{titlepage}
\begin{abstract}
Analytic-bilinear approach for construction and study of integrable
hierarchies, in particular, the KP hierarchy is discussed. It is based on the
generalized Hirota identity. This approach allows to represent generalized
hierarchies of integrable equations in a condensed form of finite functional
equations. Resolution of these functional equations leads to the $\tau $%
-function and addition formulae to it. General discrete transformations of
the $\tau $-function are presented in the determinant form. Closed one-form
and other formulae
also arise naturally within the approach proposed. Generalized KP hierarchy
written in terms of different invariants of Combescure symmetry
transformations coincides with the usual KP hierarchy and the mKP hierarchy.
\end{abstract}
\section{Introduction}
The Kadomtsev-Petviashvili (KP) equation and the whole KP
hierarchy of equations are significant parts of the theory
of integrable equations. They arise in various fields of physics
from hydrodynamics to string theory. They are also the tools to solve
several problems in mathematics from the differential geometry
of surfaces to an algebraic geometry.

The KP hierarchy has been described and studied within the framework of
different approaches.The Sato approach \cite{Sato}
(see also \cite{Jimbo}-\cite{Segal}) and the
$\dbar$-dressing method \cite{dbar1}-\cite{dbar4} are, perhaps,
the most beautiful and powerful among them. The infinite-dimensional
Grassmannian, pseudo-differential operators, Hirota bilinear
identity and $\tau$-function are the basic ingredients
of the Sato approach which is in essence algebraic.
In contrast, the $\dbar$-method, based on the nonlocal
$\dbar$-problem for the wave function, mostly uses the analytic properties
of the wave function. These two approaches look like completely different.
On the other hand, each of them has its own advantages.
So one could expect that their marriage could be rather profitable.

A bridge between the Sato approach and the $\dbar$-dressing method has been
established by the observation that the Hirota bilinear identity can be
derived from the $\dbar$-dressing method \cite{dbar4}, \cite{Carroll}.
Elements of the
approach which combines the characteristic features of both methods, namely,
the Hirota bilinear identity from the Sato approach and the analytic
properties of solutions from the $\dbar$-dressing method have been considered
in \cite{DZM}, \cite{NLS}.

This paper is devoted to the analytic-bilinear approach to integrable
hierarchies. It is based on the generalized Hirota bilinear identity for
the wave function with simple analytic properties (Cauchy-Baker-Akhiezer
(CBA) function). This approach allows us to derive generalized hierarchies
of integrable equations in terms of the CBA function
$\psi(\lambda,\mu,{\bf x})$ in a compact, finite form. Such a generalized
equations contain integrable equations in their usual form, their
modified partners and corresponding linear problems.

The generalized Hirota bilinear identity provides various functional
equations for the CBA function. The resolution of some of them leads to the
introduction of the $\tau$-function, while others give rise to the addition
theorems for the $\tau$-function. The properties of the $\tau$-function,
such as the associated closed 1-form and its global definition are also
arise in a simple manner
within the approach under consideration.
Generic discrete transformations of the CBA function and of the $\tau$-
function are presented in the determinant form. These transformations are
the generalized form of the usual iterated Darboux transformations.

The generalized KP hierarchy is presented also in the `moving'
frame depending on the parameter. This generalized KP hierarchy
written in the different `moving' frames contains the Darboux system
of equations.

In addition to the usual infinite-dimensional symmetries the generalized KP
hierarchy possess the symmetries given by the Combescure transformations.
The invariants of these symmetry transformations have the compact forms in
terms of the CBA function. The generalized KP hierarchy written in terms of
these invariants coincide with the usual KP hierarchy , mKP hierarchy and
SM-KP hierarchy.

The present paper is devoted to the one-component KP hierarchy.
The authors plan to consider multi-component KP hierarchy and
the 2-dimensional Toda lattice in subsequent papers.

The paper is organized as follows. Generalized Hirota identity
is introduced in section 2. The generalized KP hierarchy is derived in
section 3. Combescure symmetry transformations are discussed
in section 4. Generalized KP
hierarchy in the `moving' frame is considered in section 5. The $\tau$-function is introduced in section 6. The addition formulae for the $\tau $-function are also obtained here. Transformations of the CBA function and
the $\tau$-function given by the determinant formulae and the Darboux
transformations are discussed in section 7. Closed one-form variational
formulae for
the $\tau $-function are presented in section 8.
%%%%%%%%%%%%%%%%%%%%%%%%%%%%%%%%%%%%%%%%%%%%%%%%%%%%%%%%%%%%%%%%%%%%%%%
\section{Generalized Hirota identity}
The famous Hirota bilinear identity provides us condensed
and compact form of the integrable hierarchies.
Here we will derive the generalized Hirota bilinear identity
in frame of the $\dbar$-dressing method.

The $\dbar$-dressing method (see \cite{dbar1} - \cite{dbar4})
is based on the nonlocal $\dbar$-problem
of the form
\begin{eqnarray}
\bar{\partial}_{\lambda}(\chi({\bf x} ,\lambda)-\eta({\bf x},\lambda))=
%\hat {R}\chi({\bf x},\lambda)=
\int\!\!\!\int_{\bf C}\: d\mu\wedge d\bar{\mu}\chi(\mu)g^{-1}(\mu)
R(\mu,\lambda)g(\lambda))
\label{dpr} \\
(\chi({\bf x} ,\lambda)-\eta({\bf x},\lambda))_
{|\lambda|\rightarrow\infty}\rightarrow 0.\nn
%\nn \\
%\hat{R} \chi=
%,\nonumber
\end{eqnarray}
where $\lambda \in {\bf C},$  $ \bar{\partial}_{\lambda}={\partial /
\partial \bar{\lambda}}$,
$\eta({\bf x},\lambda)$ is a rational function
of $\lambda$ (normalization). In general case
the function $\chi(\lambda)$
and the kernel $R(\lambda,\mu)$ are matrix-valued
functions.

The dependence of the solution
$\chi(\lambda)$ of the problem (\ref{dpr}) on the dynamical
variables
is hidden in the function $g(\lambda)$.
Here we will consider only the case of continuous variables,
for which
$g_i=\exp(K_i x_i)$,
where $K_i(\lambda)$ are, in general, matrix meromorphic
functions.

It is assumed that the problem (\ref{dpr}) is uniquely solvable.
The $\dbar$-dressing method allows us to construct and solve
wide classes of nonlinear PDEs which correspond to the different
choice of the functions  $K_i(\lambda)$ .

Here we will assume
that the kernel $R(\lambda,\mu)$
is equal to zero in some open subset $G$  of
the complex plane
with respect to $\lambda$ and to $\mu$.
This subset should typically include all zeroes
and poles of the considered class of functions
$g(\lambda)$ and a neighborhood of infinity.

In this case the solution of the problem (1)
normalized by $\eta$
is the function \[\chi(\lambda)=
\eta({\bf x},\lambda)+\varphi({\bf x},\lambda),\]
where $\eta(\lambda)$ is a rational function
of $\lambda$ (normalization), all poles of $\eta(\lambda)$
belong to $G$,
$\varphi(\lambda)$
decreases as $\lambda \rightarrow \infty$ and is
{\em analytic} in $G$.

The special class of solutions of the $\dbar$ problem
(\ref{dpr}) normalized by $(\lambda-\mu)^{-1}$
($\eta(\lambda)=(\lambda-\mu)^{-1},\quad \mu\in G$)
is of particular importance for the whole $\dbar$-dressing method
\cite{dbar2}.
Let us consider the $\dbar$-problem (\ref{dpr}) for such solutions
and corresponding dual problem for the dual function
$\chi^{\ast}(\lambda,\mu)$:
\begin{eqnarray}
\bar{\partial}_{\lambda}\chi(\lambda,\mu)
=2\pi i \delta(\lambda-\mu)+
\int\!\!\!\int_{\bf C} d\nu\wedge d\bar{\nu}\chi(\nu,\mu)g_1(\nu)
R(\nu,\lambda)g_1(\nu)^{-1},\nn\\
\bar{\partial}_{\lambda}\chi^{\ast}(\lambda,\mu)
=-2\pi i \delta(\lambda-\mu)-
\int\!\!\!\int_{\bf C} d\nu\wedge d\bar{\nu}g_2(\nu) R(\lambda,\nu)
g_2(\nu)^{-1}\chi^{\ast}(\nu,\mu).
\end{eqnarray}
where $g_1=g(\lambda,{\bf x})$, $g_2=g(\lambda,{\bf x'})$
After simple calculations, one gets
\bea
\int\!\!\!\int_{G}d\nu\wedge d\bar{\nu}
{\partial\over\partial\bar{\nu}}\left(
\chi(\nu,\lambda;g_1)g_1(\nu)g_2(\nu)^{-1}
\chi^{\ast}(\nu,\mu;g_2)\right)=\nn\\
\int_{\partial G} \chi(\nu,\lambda;g_1)g_1(\nu)g_2(\nu)^{-1}
\chi^{\ast}(\nu,\mu;g_2)d\nu=0\,.
\label{HIROTA0}
\eea
In the
particular case $g_1=g_2$ from
(\ref{HIROTA0}) it follows
that in $\bar{G}$ the function $\chi(\lambda,\mu)$
is equal to $-\chi^{\ast}(\mu,\lambda)$
(see also \cite{ManZen}).
So finally we have
\be
\int_{\partial G} \chi(\nu,\mu;g_1)g_1(\nu)g_2^{-1}(\nu)
\chi(\lambda,\nu;g_2)d\nu=0\,.
\label{HIROTA}
\ee
Here the function $\chi(\lambda,\mu)$
possesses the following analytical properties:
$$
{\bar{\partial}}_{\lambda}\chi(\lambda,\mu)=
2\pi {\rm i}\delta(\lambda-\mu),
\quad
-{\bar{\partial}}_{\mu}
\chi(\lambda,\mu)=2\pi {\rm i}\delta(\lambda-\mu),
$$
where $\delta(\lambda-\mu)$ is a $\delta$-function,
or, in other words,
$\chi\rightarrow (\lambda-\mu)^{-1}$
as $\lambda
\rightarrow\mu$ and
$\chi(\lambda,\mu)$ is analytic function of
both variables $\lambda,\mu$ for $\lambda\neq\mu$.

In the particular case $\lambda=\mu=0$ the relation (\ref{HIROTA})
is nothing but the usual Hirota identity for the KP wave function
$\chi(\nu,0,{\bf x})$ and the dual KP wave function
$\chi^{\ast}(\nu,0,{\bf x'})$ (see e.g. \cite{Sato}-\cite{Segal}).
Just in this case
(for $\lambda=\mu=\infty$ ) the Hirota bilinear identity has been derived
from the $\dbar$-problem in \cite{Carroll}.

The identity (\ref{HIROTA}) represents itself the generalization of
the Hirota identity which is bilocal with respect to the dynamical
variables ${\bf x}$ and the spectral variables $(\lambda,\mu)$.
It can be considered as the point of departure without any reference
to the $\dbar$-dressing method. Namely, the generalized Hirota
bilinear relation (\ref{HIROTA}) is the starting point of the
analytic-bilinear approach which we will consider in this paper.
The double bilocality (with respect to ${\bf x}$ and
$(\lambda,\mu)$ provides us an additional freedom which will allow
us to represent integrable hierarchies in a unified and condensed form.

Introducing the function $\psi(\lambda,\mu,{\bf x})$
via
\be
\psi(\lambda,\mu;g)=g^{-1}(\mu)\chi(\lambda,\mu;g)g(\lambda)\,,
\label{substitution}
\ee
one gets another form of the generalized Hirota equation
\be
\int_{\partial G} \psi(\nu,\mu;g_1)
\psi(\lambda,\nu;g_2)d\nu=0\,.
\label{HIROTA1}
\ee
Note that in the framework of algebro-geometric technique
the function $\psi(\lambda,\mu)$ corresponds to the
Cauchy-Baker-Akhiezer kernel on the Riemann surface
(see \cite{Orlov}). We will refer to the function
$\psi(\lambda,\mu)$ as the Cauchy-Baker-Akhiezer (CBA)
function.

We will assume in what follows that we are able to find
solutions for the relation (\ref{HIROTA}) somehow.
In particular, it can be done by the
$\dbar$-dressing method.

The general setting of the problem
of solving (\ref{HIROTA}) requires some modification of
Segal-Wilson Grassmannian approach \cite{Segal}.
Let us consider two
linear spaces $W(g)$ and $\widetilde W(g)$
defined by the function $\chi(\lambda,\mu)$
(satisfying (\ref{HIROTA}))
via equations connected with equation (\ref{HIROTA})
\bea
\int_{\partial G} f(\nu;g)\chi(\lambda,\nu;g)
d\nu=0,
\label{W}\\
\int_{\partial G}\chi(\nu,\mu;g)h(\nu;g)
d\nu=0 ,
\label{W'}
\eea
here
$f(\lambda)\in W$, $h(\lambda)\in \widetilde W$; $f(\lambda)$,
$h(\lambda)$ are defined in $\bar G$.

It follows from the definition of
linear spaces $W,\;\widetilde W$ that
\bea
f(\lambda)&=&2\pi {\rm i}\int\!\!\!\int_G
\eta(\nu)\chi(\lambda,\nu)d\nu\wedge d\bar{\nu},\quad
\eta(\nu)=\left({\partial\over
\partial\bar\nu}
f(\nu)\right),\nn\\
h(\mu)&=-&2\pi {\rm i}\int\!\!\!
\int_G \chi(\nu,\mu)\widetilde\eta(\nu) d\nu\wedge d\bar{\nu},\quad
\widetilde\eta(\nu)=\left({\partial\over \partial\bar\nu}
h(\nu)\right).
\label{basis}
\eea
These formulae in some sense provide an expansion of the
functions $f\,,h$ in terms of the basic function $\chi(\lambda,\mu)$.
The formulae (\ref{basis}) readily imply that linear spaces
$W,\; \widetilde W$ are transversal to the space of holomorphic functions
in $G$ (transversality property).

{}From the other point of view, these formulae define a map of
the space of functions (distributions)
on $\bar G$ $\eta,\;\widetilde\eta$
to the spaces
$W$, $\widetilde W$. We will call
$\eta$ ($\widetilde\eta$) a {\em normalization} of the
corresponding function belonging to $W$ ($\widetilde W$).

The dynamics of linear spaces $W,\;\widetilde W$ looks very
simple
\be
W(g)=W_0g^{-1};\quad \widetilde W(g)=g\widetilde W_0,
\label{dynamics}
\ee
here $W_0=W(g=1)$, $\widetilde W_0=\widetilde W(g=1)$
(the formulae (\ref{dynamics})
follow from identity (\ref{HIROTA}) and the formulae (\ref{basis})).

To introduce a dependence on several variables (may be of different
type), one should consider a product of corresponding functions $g(\lambda)$
(all of them commute).

The formula (\ref{HIROTA}) is a basic tool for our construction.
Analytic properties of the CBA kernel accompanied with the
different choices of the functions $g_1$ and $g_2$ will provide
us various compact and useful relations.
%%%%%%%%%%%%%%%%%%%%%%%%%%%%%%%%%%%%%%%%%%%%%%%%%%%%%%%%%%%%%%%%%%%%%%%
\section{Generalized KP hierarchy}
In the present paper we will consider the scalar KP hierarchy.
It is generated by the generalized Hirota formula (\ref{HIROTA})
where $G$ is a unit disk
and
\be
g({\bf x},\lambda)
=\exp\left(\sum_{i=1}^{\infty}x_i\lambda^{-i}\right)\,.
\label{gKP}
\ee
Let us consider the formula (\ref{HIROTA}) with
\bea
g_1 g_2^{-1}=
g({\bf x},\nu)g^{-1}({\bf x'},\nu)
=\exp\left(\sum_{i=1}^{\infty}(x_i-x_i')\nu^{-i}\right)=
{\nu-a\over\nu-b}\,.
\label{gKP1}
\eea
where $a$ and $b$ are arbitrary complex parameters,
$a\,,b\in G$. Since $\log(1-\epsilon)
=\sum_{i=1}^{\infty}\epsilon^i/i$, one has
$$
x'_i-x_i={1\over i}a^i-{1\over i}b^i\,.
$$
Substituting the expression (\ref{gKP1}) into
(\ref{HIROTA}), one gets
\bea
&&\left({\mu-a\over\mu-b}\right)
\chi(\lambda,\mu,{\bf x}+[a])-
\left({\lambda-a\over\lambda-b}\right)
\chi(\lambda,\mu,{\bf x}+[b])+\nn\\
\nn\\
&&(b-a) \chi(\lambda,b,{\bf x}+[a])\chi(b,\mu,{\bf x}+[b])=0,
\quad \lambda\neq\mu
\label{KPbasic0}
\eea
where ${\bf x}+[a]=x_i+[a]_i\,,0 \leq i<\infty\,,[a]_i
={1\over i} a^i$.
Equation (\ref{KPbasic0}) is the simplest functional equation
for $\chi(\lambda,\mu,{\bf x})$ which follows from
the generalized Hirota equation (\ref{HIROTA}).

Residues of the l.h.s. of (\ref{KPbasic0}) at the
poles $\mu=b$ and $\lambda=b$ vanish identically.
Evaluating the l.h.s. of (\ref{KPbasic0}) at
$\mu=a$ and $\lambda=a$, one gets the equations

\bea
\left({\lambda-a\over\lambda-b}\right)
\chi(\lambda,a,{\bf x}+[b])&=&
(b-a) \chi(\lambda,b,{\bf x}+[a])\chi(b,a,{\bf x}+[b])\,,
\label{KPbasic01}\\
\left({\mu-a\over\mu-b}\right)
\chi(a,\mu,{\bf x}+[a])&=&
(a-b)\chi(a,b,{\bf x}+[a])\chi(b,\mu,{\bf x}+[b])\,,
\label{KPbasic02}\\
a\neq b\,.&&\nn
\eea
Equations (\ref{KPbasic01}) and  (\ref{KPbasic02})
imply that
\bea
&&(\lambda-\mu)\chi(\lambda,\mu,{\bf x})=\nn\\
\nn\\
&&{(\lambda-a)\chi(\lambda,a,{\bf x}-[a]+[\mu])\over
(\mu-a)\chi(\mu,a,{\bf x}-[a]+[\mu])}=
{(a-\mu)\chi(a,\mu,{\bf x}+[a]-[\lambda])\over
(a-\lambda)\chi(a,\lambda,{\bf x}+[a]-[\lambda])}\,.
\label{KPbasic03}
\eea
Since $(\mu-a)\chi(\mu,a)\rightarrow 1$ as $\mu-a\rightarrow 0$,
one gets from (\ref{KPbasic03})
\be
(\lambda-\mu)^2\chi(\lambda,\mu,{\bf x}+[\lambda])
\chi(\mu,\lambda,{\bf x}+[\mu])=-1\,.
\label{KPbasic04}
\ee
We will solve the equations
(\ref{KPbasic01}) -(\ref{KPbasic04}) in the
next section. Now, let us consider the particular form of equation
(\ref{KPbasic0}) for $b=0$. In terms of the CBA function
it reads
\be
\psi(\lambda,\mu,{\bf x}+[a])-\psi(\lambda,\mu,{\bf x})=
 a \psi(\lambda,0,{\bf x}+[a])\psi(0,\mu,{\bf x});\quad
x'_i-x_i={1\over i} a^i\,.
\label{KPbasic}
\ee
This equation is a condensed
finite form of the whole KP-mKP hierarchy.
Indeed, the expansion of this relation over $a$ generates
the KP-mKP hierarchies (and dual hierarchies)
and linear problems for them.
To demonstrate this, let us take the first three equations given by the
expansion of (\ref{KPbasic}) over~$ a$
\bea
a:&\;\;&\psi(\lambda,\mu,{\bf x})_x=
\psi(\lambda,0,{\bf x})\psi(0,\mu,{\bf x})\,,
\label{KPbasic1}\\
a^2:&\;\;&\psi(\lambda,\mu,{\bf x})_y=
\psi(\lambda,0,{\bf x})_x\psi(0,\mu,{\bf x})-
\psi(\lambda,0,{\bf x})\psi(0,\mu,{\bf x})_x\,,
\label{KPbasic2}\\
a^3:&\;\;&\psi(\lambda,\mu,{\bf x})_t=
{1\over 4}\psi(\lambda,\mu,{\bf x})_{xxx} -
{3\over 4}\psi(\lambda,0,{\bf x})_x\psi(0,\mu,{\bf x})_x+\nn\\
&&{3\over 4}\left ( \psi(\lambda,0,{\bf x})_y\psi(0,\mu,{\bf x})-
\psi(\lambda,0,{\bf x})\psi(0,\mu,{\bf x})_y\right )
\label{KPbasic3} \\
&&x=x_1;\quad y=x_2;\quad t=x_3\,.\nn
\eea
In the order $ a^2$ equation (\ref{KPbasic})
gives rise equivalently to the equations
\bea
\psi(\lambda,\mu,{\bf x})_y
-\psi(\lambda,\mu,{\bf x})_{xx}&=&-
2\psi(\lambda,0,{\bf x})\psi(0,\mu,{\bf x})_x\,,
\label{KPbasic2+}\\
\psi(\lambda,\mu,{\bf x})_y
+\psi(\lambda,\mu,{\bf x})_{xx}&=&
2\psi(\lambda,0,{\bf x})_x\psi(0,\mu,{\bf x})\,,
\label{KPbasic2-}
\eea
Evaluating the first equation at $\mu=0$, the second
at $\lambda=0$ one gets
\bea
f({\bf x})_y-f({\bf x})_{xx}&=&
u({\bf x})f({\bf x}),\label{LKP}\\
\widetilde f({\bf x})_y+\widetilde f({\bf x})_{xx}&=&-
u({\bf x})\widetilde f({\bf x})
\label{LKPd}
\eea
where $u({\bf x})=-2\psi(0,0)_x$ and
$$f=\int\psi(\lambda,0)
\rho(\lambda)d\lambda\,,$$
$$\widetilde f=\int\widetilde\rho(\mu)\psi(0,\mu)
d\mu\,;$$
$\rho(\lambda)$ and $\widetilde\rho(\mu)$ are some
arbitrary functions.

In a similar manner, one obtains from (\ref{KPbasic1})-
(\ref{KPbasic3}) the equations
\bea
f_t-f_{xxx}={3\over 2}u f_x+
{3\over 4}(u_x+\partial_x^{-1}u_y)f\,,
\label{AKP}\\
\widetilde f_t-\widetilde f_{xxx}={3\over 2}u \widetilde f_x+
{3\over 4}(u_x-\partial_x^{-1}u_y)\widetilde f\,.
\label{AKPd}
\eea
Both the linear system (\ref{LKP}), (\ref{AKP}) for the
wave function $f$ and the linear system (\ref{LKPd}), (\ref{AKPd})
for the
wave function $\widetilde f$ give rise to the same
KP equation
\be
u_t={1\over 4}u_{xxx}+{3\over2}uu_x+{3\over4}\partial_x^{-1}u_{yy}\,.
\ee
To derive linear problems for the mKP and dual mKP
equations, we integrate equations
(\ref{KPbasic1}), (\ref{KPbasic2+}), (\ref{KPbasic2-})
and (\ref{KPbasic3})
with the two arbitrary functions  $\rho(\lambda)$, $\widetilde\rho(\mu)$
\bea
\Phi({\bf x})_x&=&
f({\bf x})\widetilde f ({\bf x})\,,\\
\Phi({\bf x})_y-\Phi({\bf x})_{xx}&=&-
2f({\bf x})\widetilde f({\bf x})_x\,,\\
\Phi({\bf x})_y+\Phi({\bf x})_{xx}&=&
2f({\bf x})_x\widetilde f({\bf x})\,,\\
\Phi({\bf x})_t-\Phi({\bf x})_{xxx} &=&-
{3\over2}f({\bf x})_x\widetilde f({\bf x})_x-
{3\over4}(f({\bf x})\widetilde f({\bf x})_y-
f({\bf x})_y\widetilde f({\bf x}))
\eea
where
$$\Phi=\int\!\!\!\int
\widetilde\rho(\mu)\psi(\lambda,\mu)
\rho(\lambda)d\lambda\,d\mu\,.$$
Using the first equation to exclude $f$ from the second
(and $\widetilde f$ from the third), we obtain
\bea
\Phi_y-\Phi_{xx}&=&
v({\bf x})\Phi_x\,,
\label{mKPlinear+}\\
\Phi_y+\Phi_{xx}&=&
-\widetilde v({\bf x})\Phi_x
\label{mKPlinear-}
\eea
where $v=-2{\widetilde f({\bf x})_x\over\widetilde f({\bf x})}$,
$\widetilde v=2{f({\bf x})_x\over f({\bf x})}$.

Similarly, one gets from (\ref{KPbasic3})
\bea
\Phi_t-\Phi_{xxx}&=&
{3\over2}v({\bf x})\Phi_{xx}+
{3\over4}(v_x+v^2+\partial_x^{-1}v_y)\Phi_x\,,
\label{mKPA+}\\
\Phi_t-\Phi_{xxx}&=&
{3\over2}\widetilde v({\bf x})\Phi_{xx}+
{3\over4}(\widetilde v_x+v^2-\partial_x^{-1}\widetilde v_y)\Phi_x\,.
\label{mKPA-}
\eea

The system (\ref{mKPlinear+}), (\ref{mKPA+}) gives rise
to the mKP equation
\be
v_t=v_{xxx}+{3\over4} v^2v_x +3v_x\partial_x^{-1}v_y+
3\partial_x^{-1}v_{yy}\,,
\label{mKP}
\ee
while the system
(\ref{mKPlinear-}), (\ref{mKPA-}) leads to the dual mKP equation,
which is obtained from the (\ref{mKP}) by the substitution
$v\rightarrow\widetilde v$, $t\rightarrow -t$, $y\rightarrow -y$,
$x\rightarrow -x$.

So the function $\Phi$ is simultaneously a wave function
for the mKP and dual mKP linear problems with different potentials,
defined by the dual KP (KP) wave functions.

Using the equation (\ref{KPbasic3}) and relations (\ref{mKPlinear+})
and (\ref{mKPlinear-}), one also obtains an equation for
the function $\Phi$
\bea
\Phi_t-{1\over4}\Phi_{xxx}-{3\over8}
{\Phi_y^2-\Phi_{xx}^2\over\Phi_x}+
{3\over 4}\Phi_x W_y&=&0 ,\quad W_x={\Phi_y\over\Phi_x}\,.
\label{singman}
\eea
This equation first arose in Painleve analysis of the KP
equation as a singularity manifold equation \cite{Weiss}.

It is tedious but absolutely straightforward check that
the expansion of (\ref{KPbasic}) in higher orders of $a$
generates\\
{\bf 1}) the whole hierarchy of KP singularity manifold equations
for $\psi(\lambda,\mu)$ (or $\Phi({\bf x})$\\
{\bf 2}) the hierarchy of linear problems for the mKP and dual
mKP equations, where $\psi(\lambda,\mu)$ (or $\Phi({\bf x})$
is the common wave function and
$v=-2(\log\psi(0,\lambda,{\bf x}))_x$,
$\widetilde v=-2(\log\psi(\lambda,0,{\bf x}))_x$
are the potentials\\
{\bf 3}) mKp hierarchy for $v$ and dual mKP hierarchy for
$\widetilde v$\\
{\bf 4}) the hierarchies of KP linear problems for $\psi(\lambda,0,{\bf x})$
and dual KP linear problems for $\widetilde\psi(\lambda,0,{\bf x})$\\
and, finally\\
{\bf 5}) the KP hierarchy of equations for
$u=-2\psi(0,0)_x$.

Note also one interesting consequence of the formula
(\ref{KPbasic04})
\be
\chi(0,\lambda,{\bf x})=-{1\over\lambda^2}
\chi^{-1}(\lambda,0,{\bf x}+[\lambda])\,.
\ee

\section{KP hierarchy in the `moving frame'. Darboux equations as the
horizontal
subhierarchy}
Now let us consider the expansion of the l.h.s. of (\ref{KPbasic0})
over $\epsilon=a-b$, where $\epsilon\rightarrow 0$. In the first
order in $\epsilon$ one gets
\bea
\Delta_1(b)\psi(\lambda,\mu,{\bf x})=\psi(b,\mu,{\bf x})
\psi(\lambda,b,{\bf x})
\eea
where $$\Delta_1(b)=\sum_{n=1}^{\infty}b^{n-1}
{\partial\over\partial x_n}.$$ In the higher orders in
$\epsilon$ one obtains the hierarchy of equations of the form
(\ref{KPbasic1})-(\ref{KPbasic3}) and their higher analogues
with the substitution $\psi(\lambda,0,{\bf x})\rightarrow
\psi(\lambda,b,{\bf x})$, $\psi(0,\mu,{\bf x})\rightarrow
\psi(b,\mu,{\bf x})$ and $${\partial\over\partial x_i}
\rightarrow \Delta_i(b)=\sum_{n=i}^{\infty}
{n!\over n(n-i)!i!}
b^{n-i}{\partial\over\partial x_i}.$$
Such a substitution is in fact nothing
but the change of dynamical variables (or the coordinates on
the group of functions $g$). Indeed, it is not difficult to show
that $ \Delta_i(b)={\partial\over\partial x_i(b)}$,
where the dynamical variables $x_i(b)$ are defined by the relation
\bea
\sum_{i=1}^{\infty}{x_i(b)\over (\lambda-b)^i}=
\sum_{i=1}^{\infty}{x_i\over (\lambda)^i}\,.
\eea
It is clear that
\bea
\left [{\partial\over\partial x_i(\lambda')},
{\partial\over\partial x_i(\lambda)}\right]=
0\,.
\eea
Note one interesting property of the derivatives
${\partial\over\partial x_i(b)}$, namely
\bea
\left [{\partial\over\partial \lambda},
{\partial\over\partial x_i(\lambda)}\right]=
(i+1){\partial\over\partial x_{i+1}(\lambda)}\,.
\eea
So the operator ${\partial\over\partial \lambda}$
is a `mastersymmetry' for all vector fields
${\partial\over\partial x_i(\lambda)}$.

The expansion of equation (\ref{KPbasic0})
up to the third order in $\epsilon $ gives
the equations
\be
\frac \partial {\partial x_1(b)}\psi (\lambda ,\mu ,x(b))=\psi (b,\mu
,x(b))\psi (\lambda ,b,(x(b)),
\label{m1}
\ee

\be
\frac \partial {\partial x_2(b)}\psi (\lambda ,\mu ,x(b))=\frac \partial
{\partial x_1(b)}\psi (\lambda ,b)\cdot \psi (b,\mu )-\psi (\lambda ,b)\cdot
\frac \partial {\partial x_1(b)}\psi (b,\mu ),
\label{m2}
\ee

\bea
\frac \partial {\partial x_3(b)}\psi (\lambda ,\mu ,x(b))=\frac 14\frac{%
\partial ^3}{\partial x_1(b)^3}\psi (\lambda ,\mu )-\frac 34\frac \partial
{\partial x_1(b)}\psi (\lambda ,b)\cdot \frac \partial {\partial x_1(b)}\psi
(b,\mu )+
\nn\\
+\frac 34\left( \frac \partial {\partial x_2(b)}\psi (\lambda ,b)\cdot \psi
(b,\mu )-\psi (\lambda ,b)\cdot \frac \partial {\partial x_2(b)}\psi (b,\mu
)\right) .
\label{m3}
\eea
The analogues of equations (\ref{KPbasic2+}),
(\ref{KPbasic2-}) have the form
\be
\frac \partial
{\partial x_2(b)}\psi (\lambda ,\mu ,x(b))-\frac{\partial ^2}{%
\partial x_1(b)^2}\psi (\lambda ,\mu )+2\psi (\lambda ,b)\frac \partial
{\partial x_1(b)}\psi (b,\mu )=0,
\label{m2+}
\ee

\be
\frac \partial {\partial x_2(b)}
\psi (\lambda ,\mu ,x(b))+\frac{\partial ^2}{%
\partial x_1(b)^2}\psi (\lambda ,\mu )-2\frac \partial {\partial x_1(b)}\psi
(\lambda ,b)\cdot \psi (b,\mu )=0.
\label{m2-}
\ee

Equations (\ref{m1})-(\ref{m3}) and
higher equations again give rise to the generalized KP
hierarchy but now in coordinates $x_i(b),i=1,2,3...$. For such KP hierarchy
written in the `moving'
frame the parameter b is an arbitrary one, but fixed.

Let us consider now equations of the type (\ref{m1})
written for several values of
$b$. We denote $x_1(b_\alpha )=\xi _\alpha ,\alpha =1,2,...,n.$ Equations
(\ref{m1})
taken for $b=b_\alpha ,\lambda =b_\beta ,\mu =b_\gamma (\alpha \neq \beta
\neq \gamma ),$ look like
\be
\frac \partial {\partial \xi _\alpha }\psi _{\beta \gamma }=\psi _{\beta
\alpha }\psi _{\alpha \gamma }\,,\quad\alpha \neq \beta \neq \gamma
\label{m5}
\ee
where $\psi _{\alpha \beta }=\psi (b_\alpha ,b_{\beta ,}x).$ The system
(\ref{m5})
is just well-known system of $N^2-N$ resonantly interacting waves.

Integrating equations (\ref{m1})
over $\mu $ with the function $\rho (\mu )$ and
evaluating the result at $b=b_\alpha ,\gamma =b_\beta $, one gets
\be
\frac{\partial f_\beta }{\partial \xi _\alpha }=\psi _{\beta \alpha
}f_\alpha\,,\quad(\alpha \neq \beta )
\label{m6}
\ee
where $f_\beta =\int d\mu \psi (b_{\beta ,}\mu )\rho (\mu )$. Analogously
one gets
\be
\frac{\partial f_\beta ^{*}}{\partial \xi _\alpha }=\psi _{\alpha \beta
}f_\alpha ^{*}\,,\quad(\alpha \neq \beta )
\label{m7}
\ee
where $f_\beta ^{*}=\int d\lambda \psi (\lambda ,b_\beta )\rho ^{*}(\lambda
) $ and $\rho ^{*}(\lambda )$ is an arbitrary function. The systems
(\ref{m6})
and (\ref{m7})
are the linear problem and dual linear problem for equations (\ref{m5}),
respectively.

Expressing $\psi _{\alpha \beta }$  via 
$f_\alpha $ and $f_\alpha ^{*}$, one
gets from (\ref{m6})
and (\ref{m7}) (using (\ref{m5}))
the same system for $f_\alpha $ and $f_\alpha ^{*}$ 
\be
\frac{\partial ^2H_\alpha }{\partial \xi _\beta \partial \xi _\gamma }=\frac
1{H_\beta }\frac{\partial H_\beta }{\partial \xi _\gamma }\frac{\partial
H_\alpha }{\partial \xi _\beta }+\frac 1{H_\gamma }
\frac{\partial H_\gamma }{%
\partial \xi _\beta }\frac{\partial H_\alpha }{\partial \xi _\gamma }%
\,,\quad(\alpha \neq \beta \neq \gamma \neq \alpha ).
\label{Darboux}
\ee

The system (\ref{Darboux})
is the Darboux system which was introduced for the first time
in the differential geometry of surfaces [14] and then was rediscovered in
the matrix form within the $\partial -$dressing method in the paper [5].
Note that the Darboux equations in the variables of the type $x_1(b_\alpha )$
have appeared also in the paper
\cite{Nijhoff} within completely different approach.

One can treat the Darboux equations (\ref{Darboux})
with different n as the horizontal
subhierarchy of the whole generalized KP hierarchy. Note that equations
(\ref{m1})-(\ref{m3})
and their higher analogues give rise to the higher resonantly interacting
waves equations.

%%%%%%%%%%%%%%%%%%%%%%%%%%%%%%%%%%%%%%%%%%%%%%%%%%%%%%%%%%%%%%%%%%%%%
\section{Combescure symmetry transformations for the
generalized KP hierarchy}
Let us consider now the symmetries of the equations
derived above.
All the higher equations of the hierarchy are, as usual,
the symmetries of each member of the hierarchy. Here we will
discuss another type of symmetries.

Since $\rho(\lambda)$ and $\widetilde\rho(\mu)$ are arbitrary functions,
equation (\ref{singman}) and the hierarchy (\ref{KPbasic})
possess the symmetry transformation
$$\Phi(\rho(\lambda),\widetilde\rho(\mu))\rightarrow \Phi'=
\Phi(\rho'(\lambda),\widetilde\rho'(\mu))\,.$$
This transformation is, in fact, the transformation which
changes the normalization of the wave functions. The fact
that such transformations are connected with the so-called
Combescure transformations,known for a long time in differential
geometry, was pointed out in \cite{DZM}.

The Combescure transformation was introduced last century
within the study of the transformation properties of
surfaces (see e.g. \cite{Darboux}, \cite{GEOMA}). It is a transformation
of surface such that the tangent vector at a given point of the surface
remains parallel. The Combescure transformation is essentially
different from the well-known B\"acklund and Darboux transformations.
The Combescure transformation plays an important role in the theory of the
systems of hydrodynamical type \cite{Tsarev}. It is also of great interest
for the theory of (2+1)-dimensional integrable systems \cite{Kon2}.
Combescure symmetry transformations
are essential part of the analytic-bilinear approach.

The Combescure transformation can be
characterized in terms of the
corresponding invariants. The simplest of
these invariants for the mKP equation is just the potential of
the KP equation L-operator expressed through the
wave function
\bea
u&=&{f({\bf x})_y-f({\bf x})_{xx} \over f({\bf x})}\,,
\label{invar1}\\
u&=&{\widetilde f({\bf x})_y-\widetilde f({\bf x})_{xx}
\over \widetilde f({\bf x})}\,,
\label{invar2}
\eea
or, in terms of the solution for the mKP (dual mKP) equation
\bea
v'_y+v'_{xx}-{1\over 2}((v')^2)_x&=&v_y+v_{xx}-{1\over 2}(v^2)_x\,,
\label{invar01}\\
\widetilde v'_y-\widetilde v'_{xx}-{1\over 2}((\widetilde v')^2)_x&=&
\widetilde v_y-\widetilde v_{xx}-{1\over 2}(\widetilde v^2)_x\,.
\label{invar02}
\eea
The solutions of the mKP equations are transformed only by
a subgroup of the Combescure symmetry group corresponding
to the change of the weight function $\widetilde\rho(\mu)$
(left subgroup)
and they are invariant under the action of the subgroup
corresponding to $\rho(\lambda)$ (vice versa for the dual
mKP).

All the hierarchy of the Combescure transformation
invariants is given by the expansion over $\epsilon$
near the point ${\bf x}$ of the
relation (\ref{KPbasic}) rewritten in the form
\bea
{\partial \over \partial \epsilon}
\left ({\widetilde f({\bf x}')-\widetilde f({\bf x})\over
\epsilon \widetilde f({\bf x})}\right ) &=&
-{1\over2}{\partial \over \partial \epsilon}
\partial^{-1}_{x'} u({\bf x}'),\quad
x'_i-x_i={1\over i}\epsilon^i;\\
{\partial \over \partial \epsilon}
\left ({f({\bf x})-f({\bf x'})\over
\epsilon f({\bf x})}\right ) &=&
{1\over2}{\partial \over \partial \epsilon}
\partial^{-1}_{x'} u({\bf x}')\,,\quad
x'_i-x_i=-{1\over i}\epsilon^i.
\eea
The expansion of the left part of these relations gives the
Combescure transformation invariants in terms of the wave functions
$\widetilde f$, $f$. To express them in terms of mKP equation (dual mKP
equation) solution, one should use the formulae
\bea
v&=&-2{{\widetilde f}_x\over \widetilde f}\,,
\quad \widetilde f=\exp(-{1\over2}\partial_x^{-1}v);\\
\widetilde v&=&2{{f}_x\over f},
\quad f=\exp({1\over2}\partial_x^{-1}\widetilde v)\,.
\eea
It is also possible to consider special Combescure transformations
keeping invariant the KP equation (dual KP equation) wave functions
(i.e. solutions for the dual mKP (mKP) equations). The first invariants
of this type are
\bea
{\Phi'_x({\bf x}) \over \widetilde f'({\bf x})}
&=&{\Phi_x({\bf x}) \over \widetilde f({\bf x})}\,,\\
{\Phi'_x({\bf x}) \over f'({\bf x})}
&=&{\Phi_x({\bf x}) \over f({\bf x})}\,.
\eea
All the hierarchy of the invariants of this type is generated
by the expansion of the left part of the following relations over
$\epsilon$
\bea
\left ({\Phi({\bf x}')-\Phi({\bf x})\over
\widetilde f({\bf x})}\right ) &=&
\epsilon f({\bf x}')\,,\quad
x'_i-x_i={1\over i}\epsilon^i;\\
\left ({\Phi({\bf x})-\Phi({\bf x'})\over
f({\bf x})}\right ) &=&
\epsilon \widetilde f({\bf x}')\,,\quad
x'_i-x_i=-{1\over i}\epsilon^i\,.
\eea

Now let us consider the equation (\ref{singman}) and all the
hierarchy given by the relation (\ref{KPbasic}).
This equation admits the Combescure group of symmetry transformations
$\Phi(\rho(\lambda),\widetilde\rho(\mu))\rightarrow \Phi'=
\Phi(\rho'(\lambda),\widetilde\rho'(\mu))$
consisting of two subgroups (right and left Combescure
transformations). These subgroups have the following invariants
\be
v={\Phi_y-\Phi_{xx}\over\Phi_x}
\label{Comb+}
\ee
and
\be
\widetilde v ={\Phi_y+\Phi_{xx}\over\Phi_x}\,.
\label{Comb-}
\ee
{}From (\ref{mKPlinear+}), (\ref{mKPlinear-}) it follows that they
just obey the mKP and dual mKP equation respectively.
The invariant for the full Combescure transformation can be
obtained by the substitution of the expression for $v$ via $\Phi$
(\ref{Comb+}) to the formula (\ref{invar01}). It reads
\be
u=\partial_x^{-1}\left({\Phi_y\over\Phi_x}\right)_y-
{\Phi_{xxx}\over\Phi_x} + {\Phi_{xx}^2-\Phi_y^2\over 2 \Phi_x^2}\,.
\label{invarfull}
\ee
{}From (\ref{LKP}), (\ref{LKPd}), (\ref{invar1}), (\ref{invar2}),
(\ref{invar01}), (\ref{invar02}) it follows that $u$ solves
the KP equation.

So there is an interesting connection between equation (\ref{singman}),
mKP-dual mKP equations and KP equation. Equation (\ref{singman})
is the unifying equation. It possesses a Combescure symmetry
transformations group. After the factorization of equation
(\ref{singman}) with respect to one of the subgroups (right or left),
one gets the mKP or dual mKP equation in terms of the invariants
for the subgroup (\ref{Comb+}), (\ref{Comb-}).
The factorization of equation (\ref{singman})
with respect to the full Combescure transformations group
gives rise to the KP equation in terms of the invariant
of group (\ref{invarfull}).

In other words, the invariant of equation (\ref{singman})
under the full Combescure group is described by the KP equation,
while the invariants under the action of its right and left subgroups
are described by the mKP or dual mKP equations.

%%%%%%%%%%%%%%%%%%%%%%%%%%%%%%%%%%%%%%%%%%%%%%%%%%%%%%%%%%%%%%%%%%%%%%%%
\section{$\tau$-function and addition formulae}
Now we will analyze the functional equations
(\ref{KPbasic0})-(\ref{KPbasic04}). Equation
(\ref{KPbasic03}), evaluated at $\mu=0$
for some $a=a_0$ gives
\bea
\lambda\chi(\lambda,0,{\bf x})=
{(\lambda-a)\chi(\lambda,a_0,{\bf x}-[a_0])\over
(-a)\chi(\mu,a,{\bf x}-[a_0])}=
{Z(\lambda\,,{\bf x})\over Z(0\,,{\bf x})}
\label{KPbasic003}
\eea
where we denote $Z(\lambda\,,{\bf x})=
(\lambda-a_0)\chi(\lambda,a_0,{\bf x}-[a_0])$.
Substituting the expression (\ref{KPbasic003}) into equation
(\ref{KPbasic03}), we get
\bea
(\lambda-\mu)\chi(\lambda,\mu,{\bf x})=
{Z(\lambda\,,{\bf x}+[\mu])\over Z(\mu\,,{\bf x}+[\mu])}\,.
\label{KPbasic004}
\eea
It is easy to check that in virtue of (\ref{KPbasic004})
equation  (\ref{KPbasic01}) is satisfied identically,
while equation (\ref{KPbasic02}) takes the form
\be
R(a,\lambda)R(\lambda,b)R(b,a)=
R(a,b)R(b,\lambda)R(\lambda,a)
\label{tri}
\ee
where $R(a,b,{\bf x})=Z(a,x+[a]+[b])$.
In terms of $R(a,b)$ we have
\bea
(\lambda-\mu)\chi(\lambda,\mu,{\bf x})=
{R(\lambda,\mu,{\bf x}-[\lambda])\over R(\mu,\mu,{\bf x}-[\mu])}\,.
\label{KPbasic005}
\eea
Thus the problem of resolving equations (\ref{KPbasic01}),
(\ref{KPbasic02}) is reduced to the single functional
equation (\ref{tri}), which is of the form of the triangle
(Yang-Baxter) equation, well-known in the quantum theory of
solvable models (see e.g. \cite{triangle}).

{}From the definition of $R(a,b,{\bf x})$ it follows that it has
a certain special structure. Indeed, since $Z(a,{\bf x})=
R(a,b,{\bf x}-[a]-[b])$, one has $R(a,b,{\bf x}-[a]-[b])=R(a,0,x-[a])$.
Consequently $R(a,b,{\bf x})=R(a,0,x+[b])$.

So we should solve the triangle equation (\ref{tri})
within the class of $R$ of the form
$R(a,b,{\bf x})=\Xi_a({\bf x}+[b])$, where $\Xi_a({\bf x})$
is some function.
Taking the logarithm of both parts of (\ref{tri}),
one gets
\be
\Theta(a,\lambda)+\Theta(\lambda,b)+\Theta(b,a)=
\Theta(a,b)+\Theta(b,\lambda)+\Theta(\lambda,a)
\label{trilog}
\ee
where $\Theta=\log R$.
Representing $\Theta$ as $\Theta(a,b,{\bf x})=\Theta_+(a,b,{\bf x})+
\Theta_-(a,b,{\bf x})$, where $\Theta_+$ and $\Theta_-$ are
respectively symmetric and antisymmetric parts of $\Theta$,
one easily concludes that $\Theta_+$ solves (\ref{trilog})
identically while $\Theta_-$  satisfies the equation
\be
\Theta_-(a,\lambda)+\Theta_-(\lambda,b)+\Theta_-(b,a)=0\,.
\label{trilog1}
\ee
Taking equation (\ref{trilog1}) at $b=0$, one gets
\be
\Theta_-(a,\lambda)=\Theta_-(0,\lambda)-\Theta_-(0,a)\,.
\label{trilog2}
\ee
Then since $\Theta(a,b,{\bf x})$ (as $R(a,b,{\bf x})$)
has the form $\Theta(a,b,{\bf x})=Z_a({\bf x}+[b])$,
where $Z_a$ are some functions, it follows from
(\ref{trilog2})
that
$$
\Theta_-(a,\lambda)=Z_{0-}({\bf x}+[\lambda])-
Z_{0-}({\bf x}+[a])\,.
$$
Then for the symmetric part of $\Theta_+$ one has
$Z_{a+}({\bf x}+[b])=Z_{b+}({\bf x}+[a])$.
Taking $b=0$, one concludes that
$Z_{a+}({\bf x})=Z_{0+}({\bf x}+[a])$
So
$\Theta_+(a,b)=Z_{0+}({\bf x}+[a]+[b])$
Thus  general
solution of(\ref{trilog}) has the form
$$
\Theta(a,b,{\bf x})= Z_{0+}({\bf x}+[a]+[b])+
Z_{0-}({\bf x}+[b])-Z_{0-}({\bf x}+[a])\,.
$$
Consequently, the general solution of (\ref{tri}) for our class
of $R(a,b,{\bf x})$ reads
\be
R(a,b,{\bf x})=R_s({\bf x}+[a]+[b])
{\tau({\bf x}+[b])\over\tau({\bf x}+[a])}
\label{trilog3}
\ee
where $R_s$ and $\tau$ are arbitrary functions.
Substituting now the expression (\ref{trilog3})
into the expression (\ref{KPbasic005}), we get
\be
\chi(\lambda,\mu,{\bf x})={1\over(\lambda-\mu)}
{\tau({\bf x}-[\lambda]+[\mu])
\over \tau({\bf x})}
\label{tauform}
\ee
This formula coincides with the formula introduced
in the paper \cite{NLS} in a completely different context.
Note that in our approach the function $\tau({\bf x})$ is still
an arbitrary function giving a general solution of
the functional equations (\ref{KPbasic01}), (\ref{KPbasic02})
through the formula (\ref{tauform}).

Now we will use the general equation (\ref{KPbasic0}).
Substituting (\ref{tauform}) into
(\ref{KPbasic0}), one gets
\bea
&&(a-\mu)(\lambda-b)\tau({\bf x}+[a]+[\mu])
\tau({\bf x}+[\lambda]+[b])+\nn\\
&&(\lambda-a)(b-\mu)\tau({\bf x}+[\lambda]+[a])
\tau({\bf x}+[b]+[\mu])+\nn\\
&&(b-a)(\lambda-\mu)\tau({\bf x}+[b]+[a])
\tau({\bf x}+[\lambda]+[\mu])
=0\,.
\label{taubasic}
\eea
It is nothing but the simplest addition formula for the
$\tau$-function derived in \cite{Sato},
which is closely connected
with the Fay's trisecant formula \cite{Fay}.

Generalized Hirota identity gives rise also to other
addition formulae from \cite{Sato}.
Indeed, let us choose in (\ref{HIROTA})
\be
g({\bf x})g^{-1}({\bf x'})=\prod_{\alpha=1}^n
{\nu-a_{\alpha}\over \nu-b_{\alpha}}
\ee
where $n$ is an arbitrary integer and
${\bf x'}-{\bf x}=\sum_{\alpha=1}^n [a_{\alpha}]-[b_{\alpha}]$.
In this case equation (\ref{HIROTA}) gives
\bea
\prod_{\alpha=1}^n{\mu-a_{\alpha}\over\mu-b_{\alpha}}
\chi\left(\lambda,\mu,{\bf x}+\sum_{\alpha=1}^n[a_{\alpha}]\right)-
\prod_{\alpha=1}^n {\lambda-a_{\alpha}\over\lambda-b_{\alpha}}
\chi\left(\lambda,\mu,{\bf x}+\sum_{\alpha=1}^n[b_{\alpha}]\right)+&&\nn\\
\sum_{\alpha=1}^n
(b_{\alpha}-a_{\alpha})\prod_{\gamma\,,\gamma\neq\alpha}
{b_{\alpha}-a_{\gamma}\over b_{\alpha}-b_{\gamma}}
\chi\left(\lambda,b_{\alpha},{\bf x}+\sum_{\delta=1}^n[a_{\delta}]\right)
\chi\left(b_{\alpha},\mu,{\bf x}+\sum_{\delta=1}^n[b_{\delta}]\right)=0\,.&&
\label{KPbasic00}
\eea
Substituting the expression (\ref{tauform}) into (\ref{KPbasic00})
and shifting ${\bf x}\rightarrow {\bf x}+[\lambda]$, one gets
\bea
&&
\prod_{\alpha=1}^n{(\mu-a_{\alpha})\over(\mu-b_{\alpha})(\mu-\lambda)}
\tau\left({\bf x}+[\mu]+\sum_{\gamma=1}^n[a_{\gamma}]\right)
\tau\left({\bf x}+[\lambda]+\sum_{\gamma=1}^n[b_{\gamma}]\right)-\nn\\
&&
\prod_{\alpha=1}^n
{(\lambda-a_{\alpha})\over(\lambda-b_{\alpha})(\lambda-\mu)}
\tau\left({\bf x}+[\mu]+\sum_{\gamma=1}^n[b_{\gamma}]\right)
\tau\left({\bf x}+[\lambda]+\sum_{\gamma=1}^n[a_{\gamma}]\right)
+\nn\\
&&
\sum_{\alpha=1}^n
{(b_{\alpha}-a_{\alpha})\over(b_{\alpha}-\mu)(\lambda-b_{\alpha})}
\prod_{\gamma\,,\gamma\neq\alpha}
{b_{\alpha}-a_{\gamma}\over b_{\alpha}-b_{\gamma}}
\times\nn\\&&
\tau\left({\bf x}+[b_{\alpha}]+\sum_{\beta=1}^n[a_{\beta}]\right)
\tau\left({\bf x}+[\lambda]+[\mu]+
\sum_{\beta\,,\beta\neq\alpha}[b_{\beta}]\right)
=0\,,
\label{taubasic00}
\eea
It is not difficult to check (after some renotations) that equation
(\ref{taubasic00}) coincide with the Pl\"ucker's relations
for universal Grassmannian manifold (see Theorem 3 of \cite{Sato}).
That means, according to the Theorems 1-3 of the
paper \cite{Sato} that $\tau$ is the $\tau$ function of the
KP hierarchy. Note that the formulae (\ref{tauform}),
(\ref{taubasic00}) provide solutions simultaneously for the KP,
mKP (dual mKP) and SM-KP hierarchies.
%%%%%%%%%%%%%%%%%%%%%%%%%%%%%%%%%%%%%%%%%%%%%%%%%%%%%%%%%%%%%%%%%%%%%%%
\section{Determinant formulae for transformations.}
The analytic-bilinear approach allows to represent in a compact from not
only the integrable hierarchies but also rather general transformations
acting in the space of solutions.

We will consider here the transformations which are equivalent to the action
of an arbitrary meromorphic function $g(\lambda )$ on the CBA function $\chi
(\lambda ,\mu ,x)$. Let $g(\lambda )$ be a meromorphic function in $G$ which
has simple poles at the points $a_i\;(i=1,2,...,n)$ and simple zeros at the
points $b_i\;(i=1,2,...,n)$, i.e. $g(\lambda )=\prod_{i=1}^n\frac{\lambda -a_i}{
\lambda -b_i}$. To construct the transformed CBA function $\chi ^{\dagger
}(\lambda ,\mu )$ it sufficient to find a solution of the equation

\[
\int_{\partial G}\chi (\nu ,\mu )g(\nu )\chi ^{\dagger }(\lambda ,\nu )d\nu
=0
\]
where $\chi ^{\dagger }$ has the same normalization $((\lambda -\mu )^{-1})$
as $\chi $ .

The simplest way to find $\chi ^{\dagger }$ consists in the use the
following consequence of equation (\ref{HIROTA})
\[
\int_{\partial G}\int_{\partial G}d\nu\, d\rho \chi (\nu ,\mu )g(\nu )\chi
^{\dagger }(\rho ,\nu ,g)g^{-1}(\rho )\chi (\lambda ,\rho )=0.
\]
Using this formula, one finds
\be
\chi ^{\dagger }(\lambda ,\mu )=g^{-1}(\lambda )g(\mu )\frac{\det \Delta
_{n+1}}{\det \Delta _n}
\label{determinant}
\ee
where
\[
\Delta _{m\,,\alpha \beta }
=\chi (a_\alpha ,b_\beta ),\quad\alpha ,\beta =1,2,...,m
\]
and $a_{n+1}=\lambda$, $b_{n+1}=\mu $.

The formula (\ref{determinant})
defines the generic discrete transformation of $\chi $. In
terms of the $\tau -$function this transformation has the form
\[
\frac{\tau ^{\dagger }(x-[\lambda ]+[\mu ])}{\tau ^{\dagger }(x)}=(\lambda
-\mu )g^{-1}(\lambda )g(\mu )\frac{\det F_{n+1}}{\tau (x)\det F_n}
\]
where
\[
F_{n\,,\alpha \beta }
=\frac{\tau (x-[a_\alpha ]+[b_\beta ])}{a_\alpha -b_\beta }
\]
In the simplest case $n=1$ we have $(b_1=b\,,a_1=a)$
\be
\chi ^{\dagger }(\lambda ,\mu )=\frac{(\lambda -b)(\mu -a)}
{(\lambda -a)(\mu
-b)}\frac{
\left|
\begin{array}{cc}
\chi (a,b) & \chi (a,\mu ) \\
\chi (\lambda ,b) & \chi (\lambda ,\mu )
\end{array}
\right|
}{\chi (a,b)}
\label{determinant1}
\ee
and
\[
\frac{\tau ^{\dagger }(x-[\lambda ]+[\mu ])}{\tau ^{\dagger }(x)}=\frac{%
(\lambda -\mu )(a-b)(\lambda -b)(\mu -a)}{(\lambda -a)(\mu -b)}\frac{
\left|
\begin{array}{cc}
\frac{\tau (x-[a]+[b])}{a-b} & \frac{\tau (x-[a]+[\mu ])}{a-\mu } \\
\frac{\tau (x-[\lambda ]+[b])}{\lambda -b}
& \frac{\tau (x-[\lambda ]+[\mu ])%
}{\lambda -\mu }
\end{array}
\right|
}{\tau (x)\tau (x-[a]+[b])}
\]
where $|A|=\det A$.
One can represent the transformations
(\ref{determinant}) also in terms of the function $
\psi (\lambda ,\mu )$ . In that form
the determinant formulae (\ref{determinant}) taken at $%
\lambda =0$ or $\mu =0$ are very similar to the determinant formulae for the
iterated Darboux transformations
(see e.g. \cite{Matveev}). The formulae
(\ref{determinant}) give us the
Darboux transformations for all subhierarchies (KP, mKP, dual mKP, SM-KP
hierarchies) of the generalized KP hierarchy.

The determinant formulae  (\ref{determinant})
provide us also the multilinear relations for
for the $\tau$-function. Since $\tau^{\dagger}({\bf x})=\tau({\bf x}+
\sum_{i=1}^{n}(-[a_i]+[b_i]))$, the formula (\ref{determinant}) is the
$n$-linear relation for the $\tau$-function. It is easy to check that
in the simplest case $n=1$ this formula gives the simplest
addition formula (\ref{taubasic}). At $n=2$ the formula
looks like
\bea
&&\tau({\bf x})\tau({\bf x}-[\lambda]+[\mu]+[b_1]-[a_1]+[b_2]-[a_2])
\times
\left|
\begin{array}{cc}
{\tau({\bf x}-[a_1]+[b_1])\over a_1-b_1}&
{\tau({\bf x}-[a_1]+[b_2])\over a_1-b_2}\\
\\
{\tau({\bf x}-[a_2]+[b_1])\over a_2-b_1}&
{\tau({\bf x}-[a_2]+[b_2])\over a_2-b_2}
\end{array}
\right|=\nn\\
&&{(\lambda-\mu)(\mu-a_1)(\mu-a_2)(\lambda-b_1)(\lambda-b_2)\over
(\mu-b_1)(\mu-b_2)(\lambda-a_1)(\lambda-a_2)}\times
\nn\\
\nn\\
&&\tau({\bf x}+[b_1]-[a_1]+[b_2]-[a_2])\times
\left|
\begin{array}{ccc}
{\tau({\bf x}-[\lambda]+[\mu])\over \lambda-\mu}&
{\tau({\bf x}-[\lambda]+[b_1])\over \lambda-b_1}&
{\tau({\bf x}-[\lambda]+[b_2])\over \lambda-b_2}\\ \\
{\tau({\bf x}-[a_1]+[\mu])\over a_1-\mu}&
{\tau({\bf x}-[a_1]+[b_1])\over a_1-b_1}&
{\tau({\bf x}-[a_1]+[b_2])\over a_1-b_2}\\ \\
{\tau({\bf x}-[a_2]+[\mu])\over a_2-\mu}&
{\tau({\bf x}-[a_2]+[b_1])\over a_2-b_1}&
{\tau({\bf x}-[a_2]+[b_2])\over a_2-b_2}
\end{array}
\right|
\label{3det}
\eea
It is cumbersome but straightforward check that the relation
(\ref{3det}) is satisfied due to the higher addition formulae
(\ref{taubasic00}) with $n=2$.
%%%%%%%%%%%%%%%%%%%%%%%%%%%%%%%%%%%%%%%%%%%%%%%%%%%%%%%%%%%%%%%%%%%%%%%
\section{Global expression for the $\tau$-function and the closed one-form}
In this section we will investigate in what sense the formula
(\ref{tauform}) defines the $\tau$-function. Using this formula,
it is possible to prove that the $\tau$-function is defined
through the CBA function in terms of the closed 1-form, which
can be written both for differentials of the
variables ${\bf x}$ and
for variations of the function $g$ ($g$ is defined on the
unit circle).
Moreover, it is possible
to show that the formula (\ref{tauform}) and the definition
in terms of the closed 1-form are equivalent.

%%%%%%%%%%%%%%%%%%%%%%%%%%%%%%%%%%%%%%%%%%%%%%%%%%%%%%%%%%%%%%%%
For calculation in terms of variations, it is more
convenient to use the formula (\ref{tauform})
in the form
\be
\chi(\lambda,\mu)={\tau\left(g(\nu)\times
\left({\nu-\lambda\over\nu-\mu}\right)\right)
\over \tau(g(\nu))(\lambda-\mu)}
\label{tauform1}
\ee
(see \cite{NLS}).
Differentiating (\ref{tauform1}) with respect to $\lambda$,
one obtains
\be
-{1\over\tau(g)}\oint {\delta \tau\left(g(\nu)\times
\left({\nu-\lambda\over\nu-\mu}\right)\right)
\over\delta \log g}{1\over\nu-\lambda}d\nu
={\partial\over \partial \lambda}(\lambda-\mu)
\chi(\lambda,\mu;g)\,,
\ee
or for $\lambda=\mu$
\be
-{1\over\tau(g)}\oint {\delta \tau(g(\nu))
\over\delta \log g}{1\over\nu-\lambda}d\nu
=
\chi_r(\lambda,\lambda;g)
\ee
where $\chi_r(\lambda,\mu;g)=\chi(\lambda,\mu;g)-
(\lambda-\mu)^{-1}$.
This equation can be rewritten in the form
\be
-{1\over\tau(g)}\oint {\delta \tau(g(\nu))
\over\delta \log g}{1\over\nu-\lambda}d\nu
=\oint
\chi_r(\nu,\nu;g){1\over\nu-\lambda}d\nu\,.
\label{formvar1}
\ee
Relation (\ref{formvar1}) implies that the functionals
$-{1\over\tau(g)}{\delta \tau(g(\nu))\over\delta \log g}
$ and
$\chi_r(\nu,\nu;g)$
are identical for the class of functions analytic
outside the unit circle and decreasing at infinity.
Thus for ${\delta g\over g}$ belonging to this class
one has
\be
-\delta \log \tau(g(\nu))=
\oint
\chi_r(\nu,\nu;g){\delta g(\nu)\over g(\nu)} d\nu\,.
\label{formvar}
\ee
This expression defines a variational 1-form defining the
$\tau$-function. It is easy to prove using the identity (\ref{HIROTA})
that this form is
{\em closed} .
Indeed, according to (\ref{HIROTA})
\bea
\delta \chi(\lambda,\mu;g)=
\oint \chi(\nu,\mu;g){\delta g(\nu)\over g(\nu)}
\chi(\lambda,\nu;g)d\nu\,,\nn\\
\delta \chi_r(\lambda,\lambda;g)=
\oint\chi_r(\nu,\lambda;g){\delta g(\nu)\over g(\nu)}
\chi_r(\lambda,\nu;g)d\nu\,.
\label{variation}
\eea
So the variation
of the (\ref{formvar}) gives
\bea
-\delta^2 \log \tau(g)=
\oint\!\!\!\oint
\chi_r(\nu,\lambda;g)\chi_r(\lambda,\nu;g){\delta' g(\nu)\over g(\nu)}
{\delta g(\lambda)\over g(\lambda)} d\nu\,d\lambda\,.
\eea
The symmetry of the kernel of second variation with respect to
$\lambda$, $\nu$ implies that the form (\ref{formvar})
is closed.

So the formula (\ref{formvar}) gives the definition of the
$\tau$-function in terms of the closed 1-form. For the standard KP
coordinates
$$
{\delta g(\lambda)\over g(\lambda)}=
\sum_{n=1}^{\infty} {dx_n \over \lambda^n}\,.
$$
This formula allows us to obtain a closed 1-form in terms
of $dx_n$
\be
-\delta \log \tau(g(\nu))=
\sum_{n=0}^{\infty}
\left.{\partial^n\over\partial\nu^n}
\chi_r(\nu,\nu;g)\right|_{\nu=0}dx_n,.
\label{formvar4}
\ee
For $x=x_1$ this formula immediately gives the standard formula
$$
{\partial^2\over\partial x^2 }\log\tau={1\over 2}u
$$
where $u$ is a solution for the KP equation.

In fact it is possible to prove that the function $\tau$ defined
as the solution of the relation (\ref{formvar})
satisfies the global formula (\ref{tauform}).
To do this, we will show using the formula (\ref{formvar})
that the derivatives of
difference of logarithms
of the l.h.s. and the r.h.s.
of the expression (\ref{tauform1}) with respect to
$\lambda\,,\mu\,,\bar{\lambda}\,,\bar{\mu}$ and the
variation with respect to $g$ are equal to zero
for arbitrary $\lambda\,,\mu\,,g$.
That means
that l.h.s. and r.h.s. of (\ref{tauform}) could differ only
by the factor, and the normalization of the function $\chi$
implies that this factor is equal to 1.

First we will calculate the derivative with respect to $\lambda$
\bea
{\partial\over \partial \lambda}\left(\log
\chi(\lambda,\mu)(\lambda-\mu)-\log\tau\left(g(\nu)\times
\left({\nu-\lambda\over\nu-\mu}\right)\right)
+\log\tau(g(\nu))\right)=\nn\\
{\partial\over \partial \lambda}\log
\chi(\lambda,\mu)(\lambda-\mu)+
\oint
\chi_r\left(\nu,\nu;g(\nu'))\times
\left({\nu'-\lambda\over\nu'-\mu}\right)\right)
{1\over \lambda-\nu} d\nu=\nn\\
{\partial\over \partial \lambda}\log
\chi(\lambda,\mu)(\lambda-\mu)+
\chi_r\left(\lambda,\lambda;g(\nu)\times
\left({\nu-\lambda\over\nu-\mu}\right)\right)
\label{derivative}
\eea
The second term can be found in terms of $\chi(\lambda,\mu;g)$
using the determinant formula  (\ref{determinant1}).
The formula (\ref{determinant1}) gives
\bea
\chi(\lambda,\mu;g\times {\nu-\lambda\over\nu-\mu'})
={(\lambda-\mu')(\mu-\lambda)\over\mu-\mu'}
{\det\left(\begin{array}{cc}
\chi(\lambda,\mu;g)&\chi(\lambda,\mu';g)\\
{\partial\over \partial \lambda}\chi(\lambda,\mu;g)
&{\partial\over \partial \lambda}\chi(\lambda,\mu';g)
\end{array}\right)\over\chi(b,a;g)}\,.
\eea
Taking the regular part of this formula at $\mu=\lambda$, one
immediately obtains that the expression (\ref{derivative})
is equal to zero.

The case with the derivative over $\mu$ is analogous;
derivatives over $\bar{\lambda}$ and $\bar{\mu}$ immediately
give zero.
So now we proceed to the calculation of variation
\bea
&&\delta\left(\log
\chi(\lambda,\mu)(\lambda-\mu)-\log\tau\left(g(\nu)\times
\left({\nu-\lambda\over\nu-\mu}\right)\right)
+\log\tau(g(\nu))\right)=\nn\\
&&{1\over\chi(\lambda,\mu)}
\oint \chi(\nu,\mu;g){\delta g(\nu)\over g(\nu)}
\chi(\lambda,\nu;g)d\nu-
\oint
\chi_r(\nu,\nu;g(\nu)\times{\lambda-\nu\over \mu-\nu})
{\delta g(\nu)\over g(\nu)} d\nu+\nn\\
&&\oint
\chi(\nu,\nu;g){\delta g(\nu)\over g(\nu)} d\nu
\eea
(we have used (\ref{variation}) and (\ref{formvar})).
Using the determinant formula (\ref{determinant})
to transform the second term, one concludes
that the variation is equal to zero.

Below we will give a brief sketch of derivation of
1-form in terms of Baker-Akhiezer function.
Differentiation of (\ref{tauform}) (with shifted arguments)
with respect to $\lambda$ and $\mu$ gives
\bea
&&
\left({\partial\over\partial \lambda}+
{\partial\over\partial x_1(\lambda)}\right)
\log[(\lambda-\mu)\chi(\lambda,\mu,{\bf x})]=
-{\partial\over\partial x_1(\lambda)}\log \tau({\bf x})
\label{form1}\,,\\
&&
\left({\partial\over\partial \mu}-
{\partial\over\partial x_1(\mu)}\right)
\log[(\lambda-\mu)\chi(\lambda,\mu,{\bf x})]=
{\partial\over\partial x_1(\mu)}\log \tau({\bf x})
\label{form2}
\eea
where ${\partial\over\partial x_1(\lambda)}=
\sum_{n=1}^{\infty}\lambda^{n-1}{\partial\over\partial x_n}$.
Equation (\ref{form1}) implies that
$$
{\partial\over\partial x_n}\log \tau({\bf x})=\left .
-{1\over(n-1)!}{\partial^{n-1}\over\partial \lambda^{n-1}}
\left(\left({\partial\over\partial \lambda}+
{\partial\over\partial x_1(\lambda)}\right)
\log[(\lambda-\mu)\chi(\lambda,\mu,{\bf x})]\right)\right|_{\lambda=0}\,.
$$
Hence, we have the closed 1-form $\omega$
$$
\omega({\bf x}\,,{\bf dx})=
$$
\bea
\left .
-\sum_{n=1}^{\infty}
{1\over(n-1)!}{\partial^{n-1}\over\partial \lambda^{n-1}}
\left(\left({\partial\over\partial \lambda}+
{\partial\over\partial x_1(\lambda)}\right)
\log[(\lambda-\mu)\chi(\lambda,\mu,{\bf x})]\right)\right|_{\lambda=0}
dx_n &&
\label{form}\,.
\eea
and $\omega=d\log \tau$. Similar expression for $\omega$
can be obtained using (\ref{form2}). At $\mu=0$ the
formula  (\ref{form}) is equivalent to the formula found
in \cite{Date}.

%%%%%%%%%%%%%%%%%%%%%%%%%%%%%%%%%%%%%%%%%%%%%%%%%%%%%%%%%%%%%%%%%%%%%%%%
\subsection*{Acknowledgments}
The first author (LB) is grateful to the
Dipartimento di Fisica dell'Universit\`a
and Sezione INFN, Lecce, for hospitality and support;
(LB) also acknowledges partial support from the
Russian Foundation for Basic Research under grant
No 96-01-00841.
%\newpage
%%%%%%%%%%%%%%%%%%%%%%%%%%%%%%%%%%%%%%%%%%%%%%%%%%%%%%%%%%%%%%%%%%%%%%%%%
%%%%%%%%%%%%%%%%%%%%%%%%%%%%%%%%%%%%%%%%%%%%%%%%%%%%%%%%%%%%%%%%%%%%%%%%%

\end{document}